\documentclass[epj]{svjour}
\usepackage{latexsym}
\usepackage{graphicx}
\usepackage{amsmath}
\usepackage{color}

\begin{document}
\title{Correlations and confinement of excitations in an asymmetric Hubbard ladder}
\author{Anas Abdelwahab \and Eric Jeckelmann}
\institute{Leibniz Universit\"{a}t Hannover, Institut f\"{u}r Theoretische Physik, Appelstr.~2, 30167 Hannover, Germany}
\date{Received: date / Revised version: date}
%
\abstract{
Correlation functions and low-energy excitations are investigated in the asymmetric two-leg ladder consisting of a Hubbard chain and
a noninteracting tight-binding (Fermi) chain using the density matrix renormalization group method. 
The behavior of charge, spin and pairing correlations is discussed for the four phases found at half filling,
namely, Luttinger liquid, Kondo-Mott insulator, spin-gapped Mott insulator and correlated band insulator.
Quasi-long-range antiferromagnetic spin correlations are found in the Hubbard leg in the Luttinger liquid phase only.
Pair-density-wave correlations are studied to understand the structure of bound pairs found in the Fermi leg
of the spin-gapped Mott phase at half filling and at light doping but we find no enhanced pairing correlations.
Low-energy excitations cause variations of spin and charge densities on both legs that
demonstrate the confinement of the lowest charge excitations on the Fermi leg while the lowest spin excitations
are localized on the Hubbard leg in the three insulating phases.
The velocities of charge, spin, and single-particle excitations are investigated 
to clarify the confinement of elementary excitations in the Luttinger liquid phase.
The observed spatial separation of elementary spin and charge excitations could facilitate the coexistence
of different (quasi-)long-range orders in higher-dimensional extensions of the asymmetric Hubbard ladder.
} 
\maketitle
\section{Introduction}
\label{intro}

Asymmetric ladders with two inequivalent legs have attracted significant attention in recent years.
The one-di\-men\-sional (1D) Kondo-Heisenberg model was used to study
exotic correlations in stripe-ordered high-temperature superconductors~\cite{sik97,zac01b,ber10,dob13}
and quantum phase transitions in heavy fermions~\cite{eid11}. 
A study of pairing mechanisms in repulsive fermion systems was also based on a two-band Hubbard ladder model~\cite{alh09}.
Additionally, the effect of asymmetric couplings on exotic spin orders was investigated in a frustrated Heisenberg model~\cite{pan17}.
Finally, the stability of a Luttinger liquid coupled to an environment was examined using an asymmetric two-chain model~\cite{das01}.

The asymmetric Hubbard ladder with one Hubbard leg and one noninteracting (Fermi) leg was first proposed
to study proximity effects on antiferromagnetic spin correlations~\cite{yos09}. This study was motivated
by the coexistence of antiferromagnetism and superconducting correlations in multi-layered high-temperature 
superconductors.
Later, this model was the subject of a more systematic investigation~\cite{abd15} that uncovered a rich phase diagram at half filling,
although the model was found to be inappropriate for the primary motivation of that work
(atomic wires deposited on semiconducting substrates, see~\cite{abd17a,abd17b} for recent progress).
In particular, some features of the asymmetric Hubbard ladder resemble those of the Kondo-Heisenberg model~\cite{sik97,zac01b,ber10,dob13} and 
the symmetric two-leg Hubbard model~\cite{noa94,giamarchi}.  

Our first investigation~\cite{abd15} 
focused on the analysis of limiting cases as well as the calculation of physical properties such as excitation gaps, density profiles and spectral functions.
In the present paper, we discuss correlation functions corresponding to various types of symmetry-breaking orders
such as spin density waves (SDW) or pair density waves (PDW). These correlation functions were calculated numerically
using the density-matrix renormalization group (DMRG) method~\cite{dmrg,dmrg2,dmrg3}.
Naturally, (spontaneous) long-range order is not possible in the 1D model discussed here.
Nevertheless, the coexistence and competition between quasi-long range orders or enhanced fluctuations
in two-leg asymmetric Hubbard ladders may yield useful knowledge about the long-range orders induced by 
proximity effects that play a role in two-dimensional layered systems~\cite{yos09}.

In addition, we will investigate the distributions of charge and spin on the Hubbard and Fermi legs for low-energy excitations
by varying the number of electrons of each spin. Finally, we will discuss the spin and charge velocities of elementary excitations
in the Luttinger liquid. These data allow us to understand the spatial separation of elementary charge and spin excitations in the
asymmetric Hubbard ladder, in particular in the Luttinger liquid.
This spatial separation could facilitate the coexistence of various (quasi-)long-range orders for spin and charge 
in the Hubbard and Fermi subsystem of (quasi-)two-dimensional extensions of the asymmetric Hubbard ladder.

\section{Model and method}
\label{sec:model_method}

\subsection{Model}
\label{subsec:model}

\begin{figure}
\includegraphics[width=0.48\textwidth]{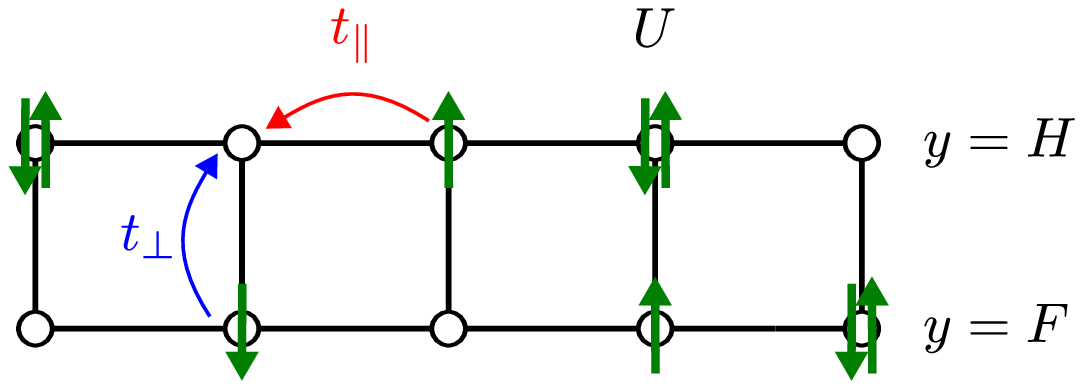}
\includegraphics[width=0.45\textwidth]{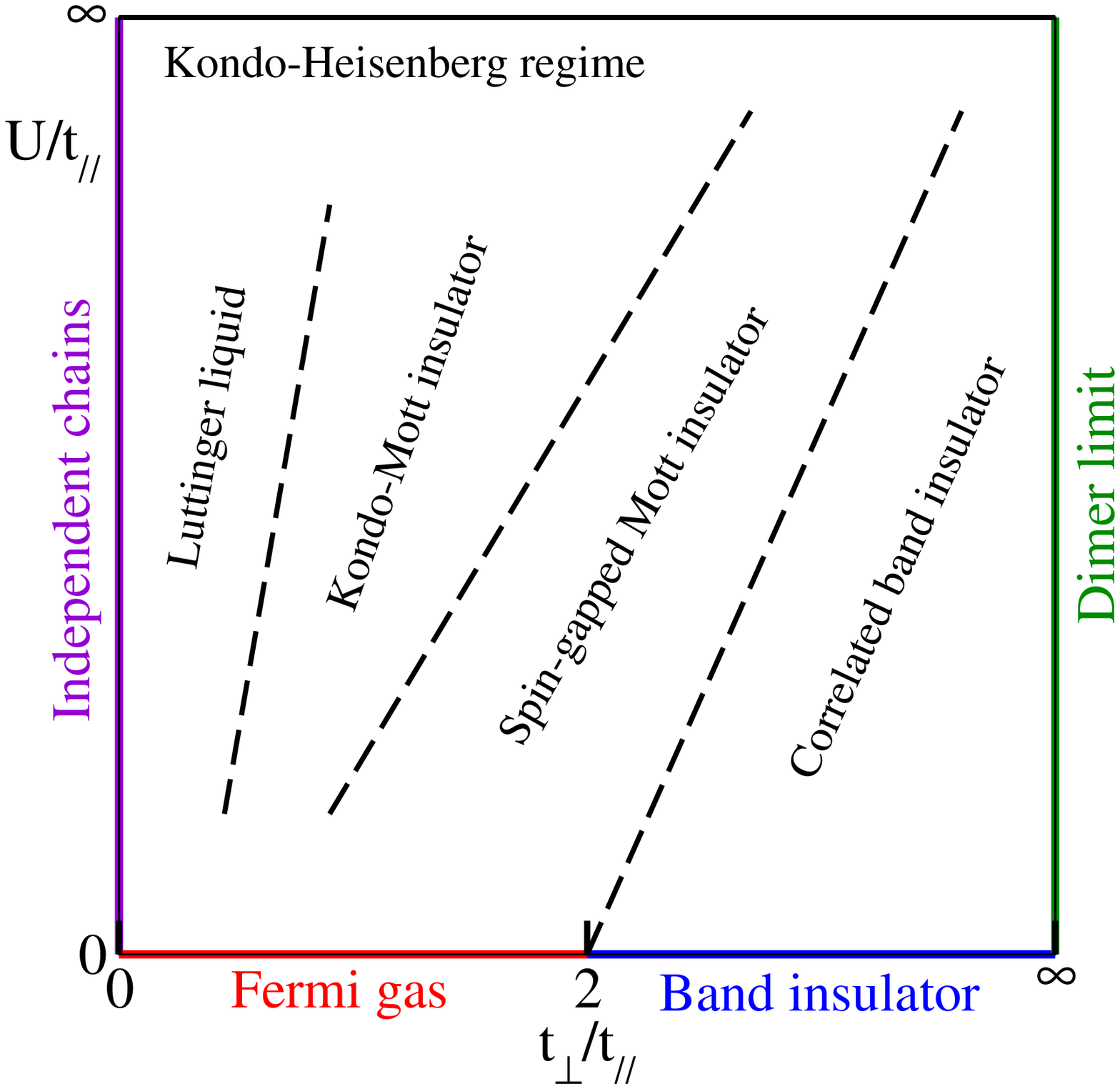}
\caption{Top figure: Sketch of the asymmetric Hubbard ladder model~(\ref{eq:hamiltonian}) with intra-leg hopping $t_{\parallel}$
and inter-leg hopping $t_{\perp}$. Electrons do not interact on the lower (Fermi, $y = F$) leg
but experience an onsite repulsion $U$ on the upper (Hubbard, $y = H$) leg.
Bottom figure: Schematic phase diagram of the half-filled asymmetric Hubbard ladder
with three insulating phases, a Luttinger liquid phase and various limiting cases identified in Ref.~\cite{abd15}.
}
\label{fig:model}
\end{figure}

The asymmetric Hubbard ladder model consists in one Hubbard leg ($y=H$) described by a Hubbard chain~\cite{hubbard-book} 
and one Fermi leg ($y=F$) described by a tight-binding chain. The two legs are connected by a single-particle hopping between adjacent sites.
The model is sketched in Fig.~\ref{fig:model}. Its Hamiltonian is 
\begin{eqnarray}
H=
&-&t_{\parallel}\sum_{x,y,\sigma} \left ( c_{x+1,y,\sigma}^{\dagger}c_{x,y,\sigma}^{\phantom{\dagger}} 
+ c_{x,y,\sigma}^{\dagger}c_{x+1,y,\sigma}^{\phantom{\dagger}} \right )  \nonumber \\
&-&t_{\perp}\sum_{x,\sigma} \left ( c_{x,\mathrm{F},\sigma}^{\dagger} c_{x,\mathrm{H},\sigma}^{\phantom{\dagger}}  +
c_{x,\mathrm{H},\sigma}^{\dagger}c_{x,\mathrm{F},\sigma}^{\phantom{\dagger}} \right ) \nonumber \\
&+&U \sum_{x} \left ( n_{x,\mathrm{H},\uparrow}-\frac{1}{2} \right ) \left (
n_{x,\mathrm{H},\downarrow}-\frac{1}{2} \right )\,.
\label{eq:hamiltonian}
\end{eqnarray}
The parameters $t_{\parallel}$ and $t_{\perp}$ describe the nearest-neighbor intra-leg and inter-leg hoppings, respectively.
The strength of the electron-electron repulsion is denoted $U$.
The operators 
$c_{x,y,\sigma}$($c^{\dagger}_{x,y,\sigma}$) annihilate (create) an electron with spin $\sigma$ on site $(x,y)$
while $n_{x,y,\sigma} = c_{x,y,\sigma}^{\dagger} c_{x,y,\sigma}^{\phantom{\dagger}}$ denote the electron number operators.
The rung index $x$ runs from $1$ to the ladder length $L$. 
The Hamiltonian is particle-hole symmetric, i.e. invariant under the transformation $c_{x,y,\sigma} \rightarrow (-1)^x c^{\dagger}_{x,y,\sigma}$.
Therefore, a half-filled ladder corresponds to $N=2L$ electrons and its Fermi energy is always equal to 0.
Moreover, it is sufficient to consider doping with additional electrons only, i.e. $N\geq 2L$.
In the singlet ground state, the numbers electrons with up and down spins are given by $N_{\uparrow}=N_{\downarrow}=\frac{N}{2}$.
We set the energy unit using $t_{\parallel}=1$.

In Ref.\cite{abd15} four different phases of the half-filled Hubbard ladder were found
for varying electron-electron coupling $U$ and inter-leg hopping $t_{\perp}$. 
The schematic phase diagram is shown in Fig.~\ref{fig:model}.
These phases are distinguished by the wave number of their low-energy excitations,
see~\cite{abd15} for details. Here we just summarize their main features.  

The first phase (starting from the right-hand side of Fig.~\ref{fig:model})
is a correlated band insulator for large inter-leg hopping $t_{\perp}$. 
(The boundary is $t_{\perp} = 2t_{\parallel}$ for $U\rightarrow 0$.)
This phase is characterized by charge and spin gaps approaching the same values and increasing linearly with $t_{\perp}$. 
The second phase is a spin-gapped Mott insulator characterized by 
finite but different charge, spin and single particle gaps at intermediate values of $U$ and $t_{\perp}$. 
The single-particle gap is larger than the charge gap resulting in a finite pair-binding energy of the order of the spin gap. 
The gaps are non-monotonic functions of the inter-leg hopping $t_{\perp}$ and the lowest excitations have incommensurate
wave number in this phase only. 
These first two phases exhibit some similarities with those observed in the symmetric Hubbard ladder~\cite{noa94,giamarchi}. 

The third phase, called a Kondo-Mott insulator, is found for large repulsive interaction $U$ and weak to intermediate inter-leg hopping
$t_{\perp}$.
It is similar (but not equivalent) to the ground state of the Kondo-Heisenberg model with charge and spin gaps
induced by the effective exchange coupling $J \sim t_{\perp}^2/U$ on a rung.
The last phase is a Luttinger liquid at weak to intermediate electron-electron repulsion
$U$ and weak inter-leg hoping $t_{\perp}$. It exhibits gapless charge and spin excitations 
with different velocities, a characteristic feature of
the dynamical separation between charge and spin in Luttinger liquids~\cite{giamarchi}.

\subsection{Method}
\label{subsec:method}

The Hamiltonian~(\ref{eq:hamiltonian}) is not exactly solvable and field-theoretical methods
have not yield much information about asymmetric ladders so far~\cite{das01,giamarchi}.
However, two-leg ladders have been studied for more than two decades with great success using DMRG methods~\cite{dmrg,dmrg2,dmrg3}.
DMRG is the most powerful numerical method for 1D dimensional correlated electron systems with short interactions. 
In this work, the finite-system DMRG was used to
calculate the ground-state properties of Hamiltonian~(\ref{eq:hamiltonian}) as described in~\cite{abd15}.
The calculations were performed on ladders with open boundary conditions and up
to $L=200$ rungs. We kept up to $m=3072$ density-matrix eigenstates to reach discarded weights smaller than $10^{-6}$.
Moreover, we extrapolated the ground-states energies to the limit of vanishing discarded weights by varying the
number of density-matrix eigenstates. 

We investigated various correlation functions of the asymmetric Hubbard ladder model~(\ref{eq:hamiltonian}) using
DMRG.
The DMRG method has often been used to investigate static correlation functions of ladder systems~\cite{noa94,dmrg2,dmrg3,noa95,noa97,rob12}.
Typically, we can obtain accurate correlation functions for finite system lengths $L$ or for short distances $x$ in infinite systems.
Consequently, the asymptotic behavior of correlations must be inferred from the short-range data using \textit{a priori} knowledge or hypotheses
about the system properties.
Despite the lower accuracy of DMRG for correlation functions and local densities than for energies,
truncation errors are negligible for the results presented here unless otherwise mentioned.
Uncertainties are mostly due to finite size and open boundary effects.

\section{Correlation functions}\label{sec:corr}

In this section we discuss the ground-state correlation functions calculated with DMRG
for the four phases found in our analysis of low-energy excitation properties~\cite{abd15}.
The charge density operator
\begin{equation}
N(x,y) = n_{x,y,\uparrow}+n_{x,y,\downarrow}
\end{equation}
is used to define the density-density correlation function
\begin{eqnarray}\label{eq:denscorr}
 C_{\text{c}}^{\alpha}(x) & = & 
\left \langle N(x_0,y)N(x_0+x,y^{\prime}) \right \rangle \nonumber \\
& - & \left \langle N(x_0,y)\right \rangle \left \langle N(x_0+x,y^{\prime})\right \rangle .
\end{eqnarray}
Intra-leg correlations corresponds to $\alpha=y=y^{\prime}=F$ for the Fermi leg
and $\alpha=y=y^{\prime}=H$ for the Hubbard leg while inter-leg correlations ($\alpha = \perp$)
are given by setting $y \neq y^{\prime}$. 
Here, we will discuss intra-leg correlations only because inter-leg correlations are always
weaker.  
The correlation functions are calculated from the middle of the ladder, $x_0= \frac{L}{2}$,
so that open boundary effects affect the results for large distances $x$ only.
Similarly, the spin density operator   
\begin{equation}
S(x,y) = n_{x,y,\uparrow}-n_{x,y,\downarrow}
\end{equation}
is used to define the spin-spin correlation function
\begin{equation}\label{eq:szszcorr}
 C_{\text{s}}^{\alpha}(x)=
\left \langle S(x_0,y)S(x_0+x,y^{\prime}) \right \rangle .
\end{equation}

\begin{figure}
\includegraphics[width=0.45\textwidth]{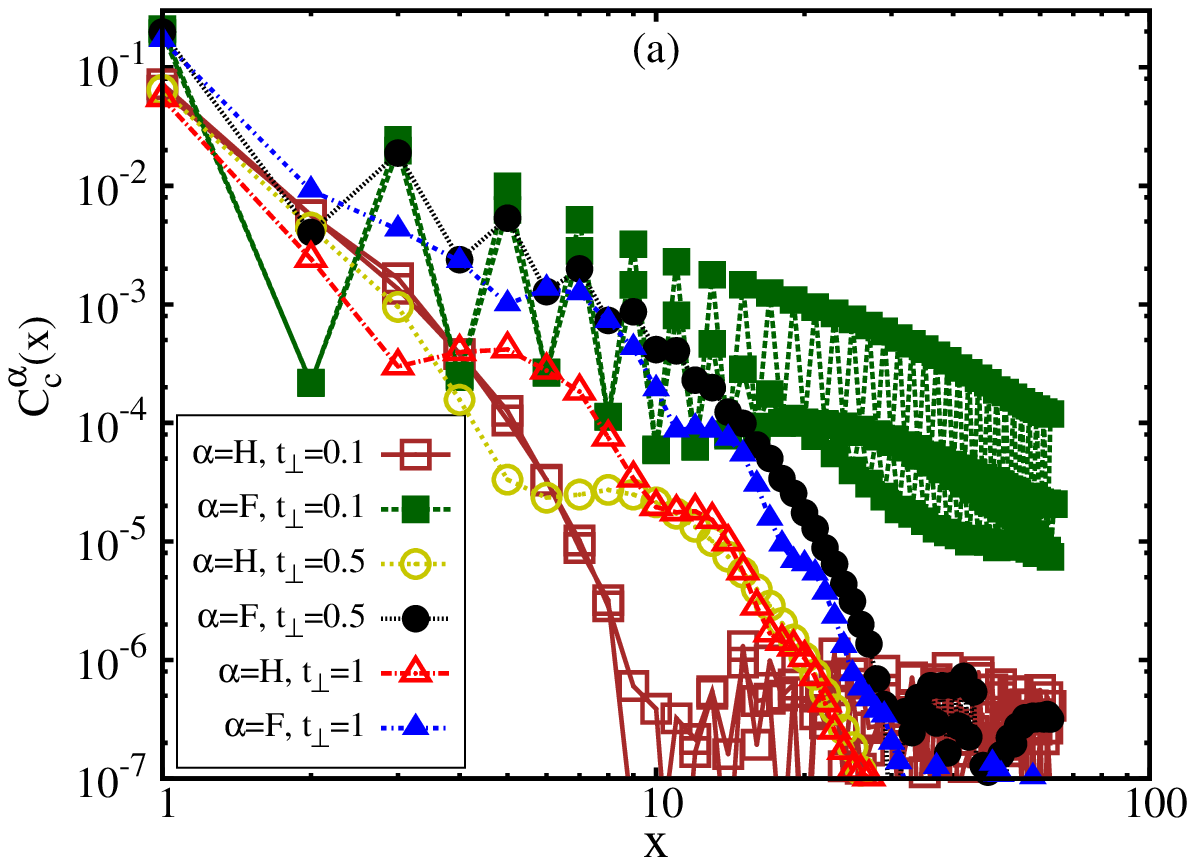}
\includegraphics[width=0.45\textwidth]{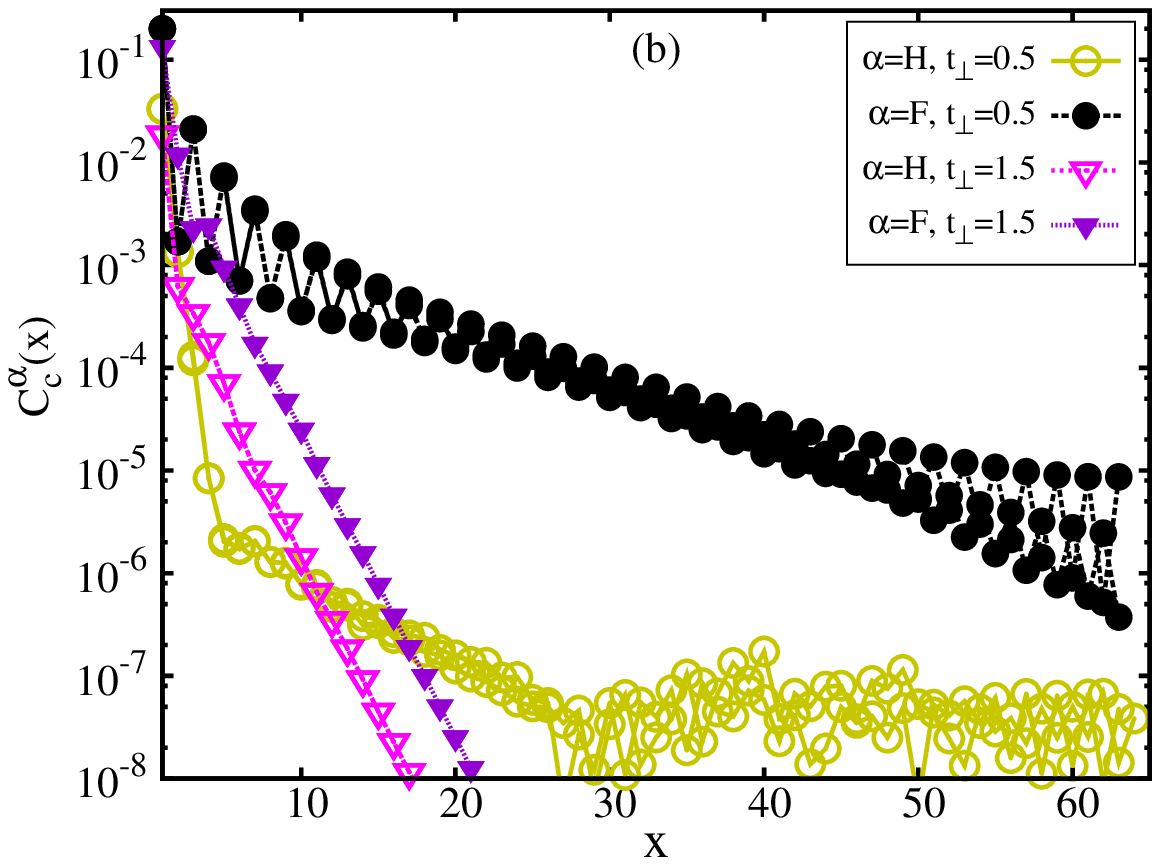}
\caption{Absolute values of the intra-leg density-density correlations~(\ref{eq:denscorr})
in the asymmetric half-filled Hubbard ladder at (a) $U=5$
and (b) $U=8$. 
The rung hopping values $t_{\perp}$ are indicated in the figures.
Open symbols show correlations in the Hubbard leg and filled symbols in the Fermi leg.}
\label{fig:denscor}
\end{figure}

Figure~\ref{fig:denscor} illustrates the density-density correlations for $U=5$ and $U=8$.
Note that we use a double logarithmic scale in Fig.~\ref{fig:denscor}(a) but Fig.~\ref{fig:denscor}(b) is a semilogarithmic plot.
For $U=5$ and $t_{\perp}=0.1$ the system is in the Luttinger liquid phase. Figure~\ref{fig:denscor}(a) shows
a power-law decay in the Fermi leg but an exponential decay in the Hubbard leg. 
The power-law behavior is expected for Luttinger liquid with gapless charge excitations
while an exponential decay is expected for the 1D half-filled Hubbard model. 
This result confirms that low-energy charge fluctuations are localized on the Fermi leg in this phase.
For the parameter sets ($U=5$, $t_{\perp}=0.5$) [in Fig. ~\ref{fig:denscor}(a)] and 
($U=8$, $t_{\perp}=0.5$) [in Fig. ~\ref{fig:denscor}(b)],
the ladder is in the Kondo-Mott phase and charge correlations decrease exponentially in both legs.
Similarly to the findings for the half-filled Kondo-Heisenberg model~\cite{ber10}, 
a charge gap is induced in the Fermi leg by the effective exchange coupling $J \sim t_{\perp}^2/U$ between both legs.
For stronger $U$ the decay becomes faster in the Hubbard leg but slower in the Fermi leg because the Mott gap 
increases with $U$ but the effective exchange coupling decreases.

Density-density correlations decay exponentially in the spin-gaped Mott phase
with slightly weaker amplitudes in the Hubbard leg as seen in Fig.~\ref{fig:denscor}(a) 
for 
$U=5$ and $t_{\perp}=1$ and in Fig.~\ref{fig:denscor}(b) for $U=8$ and $t_{\perp}=1.5$.
Finally, in the correlated band insulator (not shown) these correlations decay exponentially and similarly
fast in both legs, as expected.
DMRG truncation and convergence errors are responsible for the saturation  (i.e., the apparent long-range correlations) observed in some cases in Fig.~\ref{fig:denscor}
for large distances $x$ when  $C_{\text{c}}^{\alpha}(x)  \approx 10^{-6} - 10^{-8}$.

\begin{figure}
\includegraphics[width=0.45\textwidth]{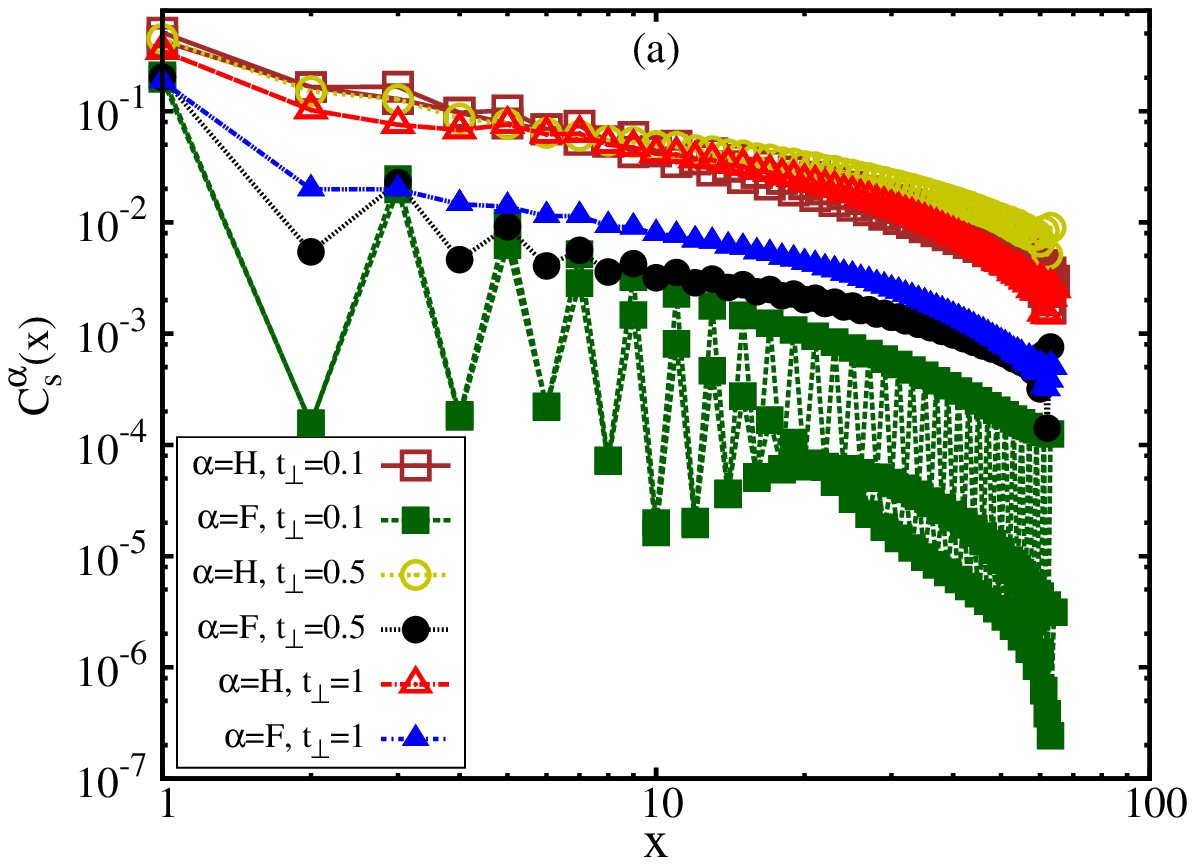}
\includegraphics[width=0.45\textwidth]{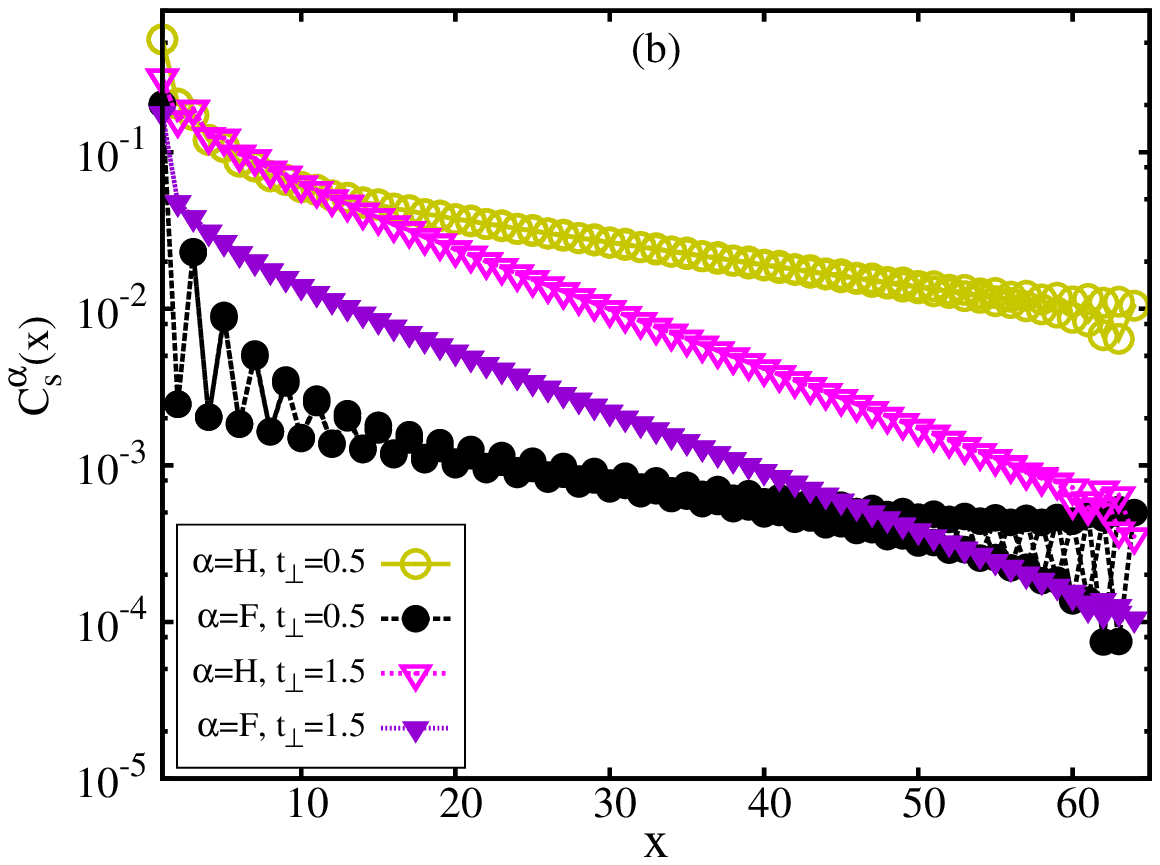}
\caption{Absolute values of the intra-leg spin correlations~(\ref{eq:szszcorr})
in the asymmetric half-filled Hubbard ladder at (a) $U=5$
and (b) $U=8$. 
The rung hopping values $t_{\perp}$ are indicated in the figures.
Open symbols show correlations in the Hubbard leg and filled symbols in the Fermi leg.}
\label{fig:szszcor}
\end{figure}

The strong antiferromagnetic correlations of the Hubbard chain induce antiferromagnetic correlations 
in the Fermi leg for $t_{\perp}\neq 0$~\cite{yos09}.
Spin correlation functions are depicted in Fig.~\ref{fig:szszcor} for
the same model parameters as used for the charge correlations in Fig.~\ref{fig:denscor}.
For the Luttinger liquid phase ($U=5$, $t_{\perp}=0.1$),
Fig.~\ref{fig:szszcor}(a) shows that the spin correlation function decays with a power-law with exponent -1 in the Hubbard leg. 
Thus the behavior of the charge and spin correlations in the Hubbard leg resembles that
of the 1D Hubbard model~\cite{hubbard-book}.    
The spin correlations in the Fermi leg follow a faster power law than in the Hubbard leg,
quite close to the one found for the density-density correlations.
This similarity between spin and charge fluctuations suggests that the Luttinger liquid in the Fermi leg 
is only weakly correlated.

The spin-spin correlations in the Kondo-Mott  phase [shown for ($U=5$, $t_{\perp}=0.5$) in Fig.~\ref{fig:szszcor}(a)
and ($U=8$, $t_{\perp}=0.5$) in Fig.~\ref{fig:szszcor}(b)] 
are weaker in the Fermi leg  than in the Hubbard leg but decay at the same rate with an apparent power law in both legs. 
Actually, they seem to be as strong as in the gapless Luttinger liquid phase.
This contradicts the observation of a finite spin gap in our previous study~\cite{abd15}. 
The existence of this gap agrees with previous findings in the half-filled Kondo-Heisenberg model~\cite{ber10}, however, 
and results from the effective exchange coupling $J \sim t_{\perp}^2/U$ between both legs.
Thus this apparent power-low behavior can only be explained by the small value of the spin gap,
resulting in correlation lengths larger than the ladder size that we can simulate with DMRG.

The spin correlations of the half-filled asymmetric Hubbard ladder were studied previously in~\cite{yos09} for
inter-leg hoppings $t_{\perp}$ corresponding to the Kondo-Mott phase. 
An apparent power-law decay was also observed (for smaller ladder sizes than in the present study)
leading to the erroneous conclusion that the system must be gapless.
The main finding in Ref.~\cite{yos09} was a non-monotonic behavior of
the induced antiferromagnetic correlations in the Fermi leg with increasing $U$. 
Our work confirms this finding and explains it as the result
of the competition between the increasing antiferromagnetic correlations in the Hubbard leg
and the decrease of the effective rung exchange coupling $J \sim t_{\perp}^2/U$ in the Kondo-Mott insulator.

A similar problem with apparent power-law SDW correlations occur in the spin-gapped Mott phase for ($U=5$, $t_{\perp}=1$), see Fig.~\ref{fig:szszcor}(a),
but for ($U=8$, $t_{\perp}=1.5$) Fig.~\ref{fig:szszcor}(b) shows clearly that the spin-spin correlations decay
exponentially.
Note that the spin gap is much smaller in this phase of the asymmetric two-leg Hubbard ladder~\cite{abd15} than in the symmetric one~\cite{noa94}
for similar parameters $U$ and $t_{\perp}$.
Consequently, it is more difficult to examine the asymptotic behavior of spin correlations.
 In the correlated band insulator (not shown), spin-spin correlation functions always decay 
exponentially fast.

The spin-gapped Mott phase is characterized by a finite pair-binding energy, which is comparable in size 
to the spin gap \cite{abd15}. Furthermore, the spin and charge density profiles show that added electrons (or holes) tend 
stay close together like a bound pair on the Fermi leg. Both features persist if the ladder is lightly doped, e.g. for
four added electrons in a $2\times128$-site ladder.
In contrast, the three other phases do not exhibit any sign of pairing. In particular,
the density profiles show that added particles tend to stay away from one another as expected for identical fermions.

We have calculated various correlation functions to investigate the nature of this pairing.
The features observed in the (lightly doped) spin-gapped Mott phase are reminiscent of the pairing 
tendency observed in the symmetric Hubbard ladder~\cite{noa94,jec98}.
There, the pairing is related to the so-called $d$-wave correlations~\cite{noa95,noa97}. 
This notion of $d$-wave order parameter is not meaningful on an asymmetric Hubbard ladder, however.

Actually, we have not found any enhanced pairing correlation in this model  
and we do not understand the nature of the observed pair binding.
Among all the pairing correlations that we have examined (singlet and triplet PDW, on-site pairs, doublon-doublon, \dots),
singlet PDW correlations decrease most slowly. 
The singlet PDW order parameter is defined as
\begin{equation}
 \Delta^{\dag}(x,y)=\frac{1}{2}
 \left( c_{x,y,\uparrow}^{\dagger}c_{x+1,y,\downarrow}^{\dagger}
 - c_{x,y,\downarrow}^{\dagger}c_{x+1,y,\uparrow}^{\dagger} \right).
\end{equation}
Thus, the (intra-leg) PDW correlation function takes the form
\begin{equation}\label{eq:pdwcorr}
 C_{\text{PDW}}^{\alpha}(x)=
\left \langle \Delta^{\dag}(x_0,y)\Delta^{\phantom{\dag}}(x_0+x,y^{\prime})
\right \rangle 
\end{equation}
where the notation is similar to (\ref{eq:denscorr}). 
It was reported that the 1D Kondo-Heisenberg model away from half filling exhibits
a spin-gapped phase with dominant PDW correlations~\cite{ber10}.
This quasi-long-range order could be the 1D precursor to striped-order in high-temperature superconductors
and has attracted much attention in recent years~\cite{ber10,dob13,rob12,jae12}.

\begin{figure}
\includegraphics[width=0.45\textwidth]{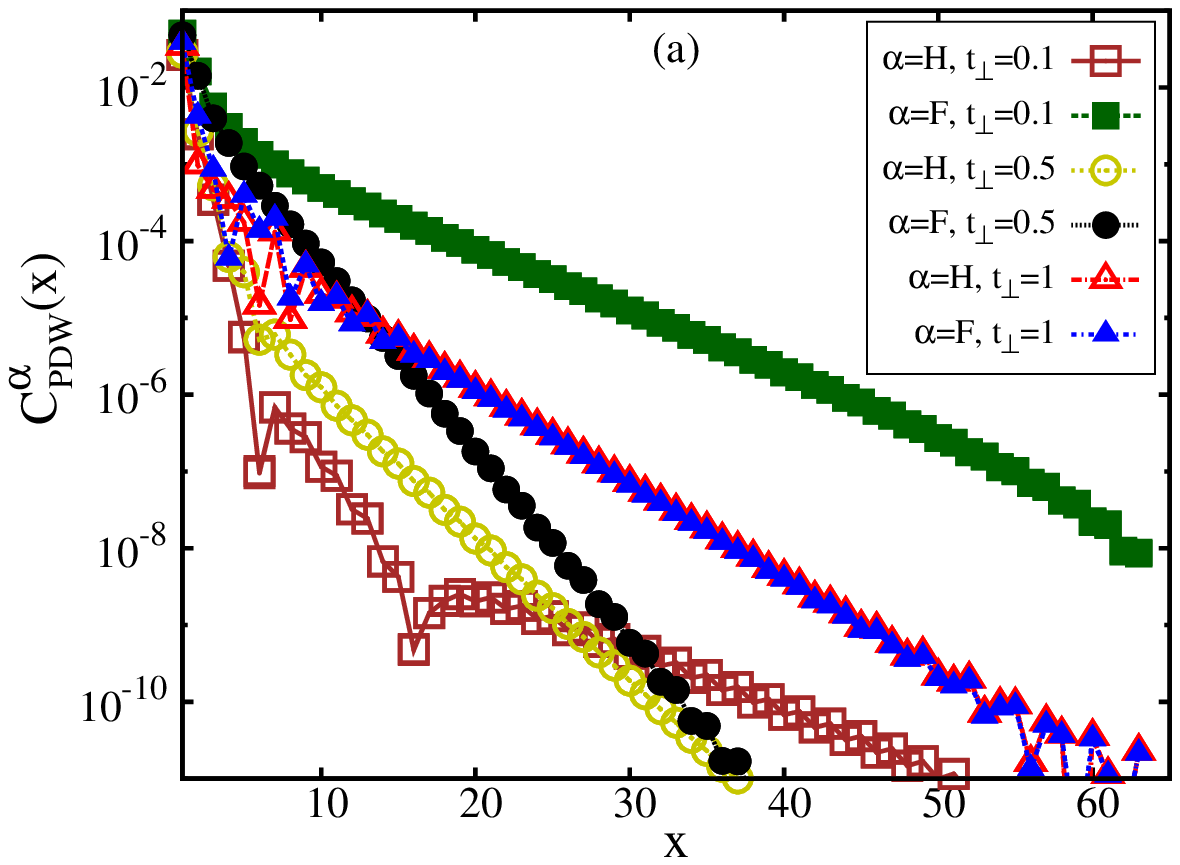}
\includegraphics[width=0.45\textwidth]{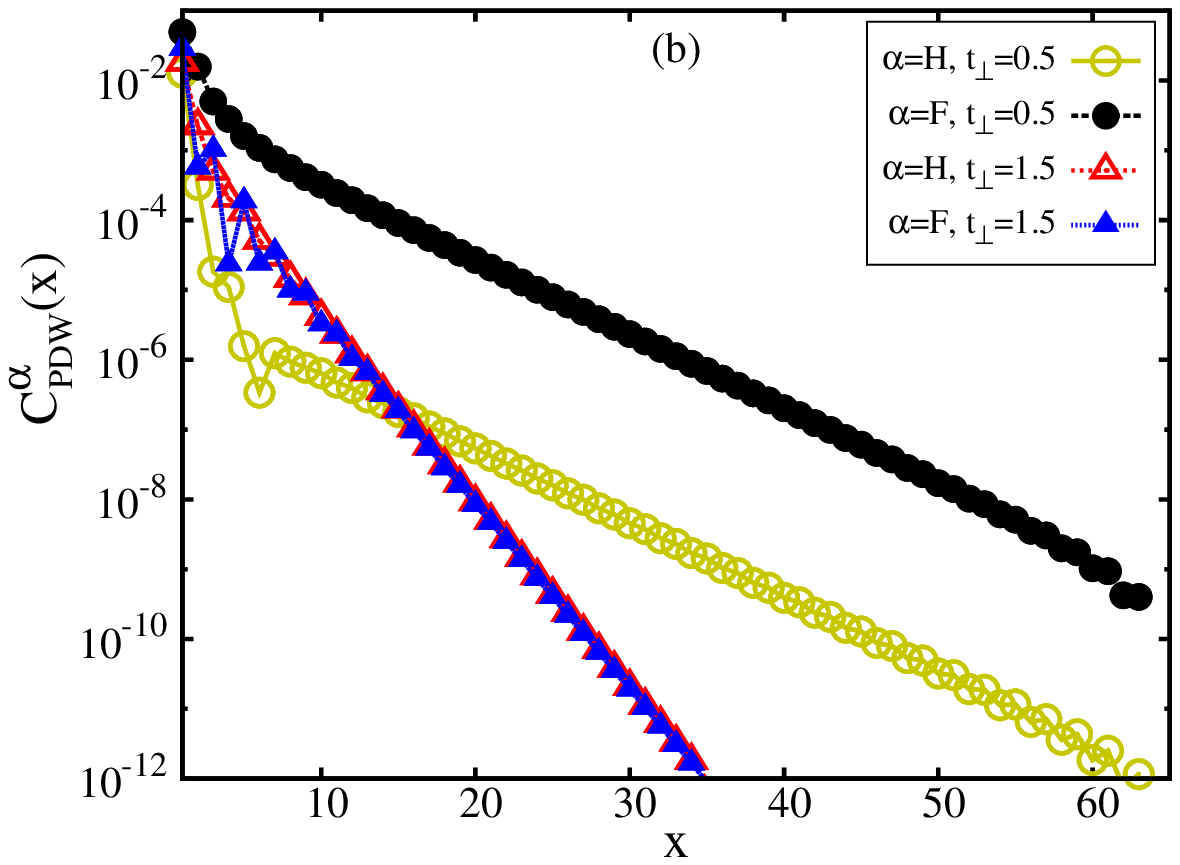}
\caption{Absolute value of the intra-leg pair-density-wave correlation functions~(\ref{eq:pdwcorr})
in the half-filled asymmetric Hubbard ladder at (a) $U=5$
and (b) $U=8$. 
The rung hopping values $t_{\perp}$ are indicated in the figures.
Open symbols show correlations in the Hubbard leg and filled symbols in the Fermi leg.}
\label{fig:pdwcor}
\end{figure}

Because of the 
similarity between the Kondo-Heisenberg model and  
the asymmetric Hubbard ladder in the strong coupling regime~\cite{abd15},
we expected to find enhanced PDW correlations in the (lightly) doped ladder~(\ref{eq:hamiltonian}).
In the Kondo-Mott phase, we have not find any sign of pairing, however. Actually, the pair binding energy
vanishes in this phase.
We think that the major reason for this discrepancy 
is that the strong rung exchange coupling (i.e., of the order of $t_{\parallel}$), for which enhanced PDW correlations were
found in the Kondo-Heisenberg model~\cite{ber10},
cannot be realized in the Kondo-Mott phase of the asymmetric Hubbard ladder, where typically $J \sim t_{\perp}^2/U \ll t_{\parallel}$.

Figure~\ref{fig:pdwcor} shows the PDW correlations for the same model parameters as in Figs.~\ref{fig:denscor} 
and~\ref{fig:szszcor}. We found that all PDW correlations decay exponentially at half filling. 
The strongest PDW correlations occur in the Fermi leg for the Luttinger liquid phase, as shown for ($U=5$, $t_{\perp}=0.1$)
in Fig.~\ref{fig:pdwcor}(a). In the Kondo-Mott phase PDW correlations are also weaker in the 
Hubbard leg than in the Fermi leg [see ($U=5$, $t_{\perp}=0.5$) in Fig.~\ref{fig:pdwcor}(a)
and  ($U=8$, $t_{\perp}=0.5$) in Fig.~\ref{fig:pdwcor}(b)] but decrease at the same rate for large distances $x$. 
In contrast, in the spin-gapped Mott insulator [($U=5$, $t_{\perp}=1$) in Fig.~\ref{fig:pdwcor}(a)
and ($U=8$, $t_{\perp}=1.5$) in Fig.~\ref{fig:pdwcor}(b)] and in the correlated band insulator (not shown)
PDW correlations are almost equal in both legs.  

\begin{figure}
\includegraphics[width=0.45\textwidth]{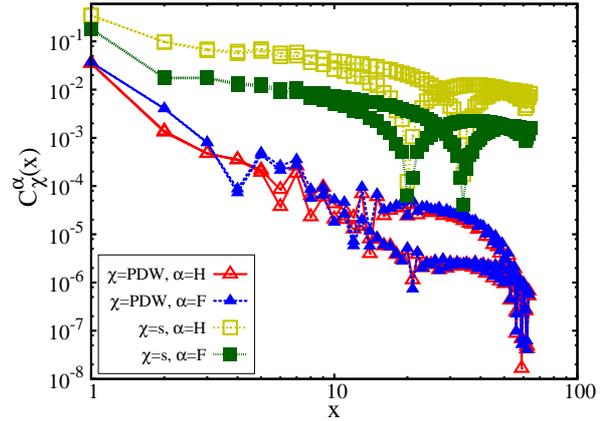}
\caption{Absolute value of the intra-leg SDW and PDW correlation functions~(\ref{eq:szszcorr}) and~(\ref{eq:pdwcorr})
in the asymmetric Hubbard ladder doped away from half filling by adding 4 electrons.
The model parameters are  $U=5$ and $t_{\perp}=1$.
Open symbols indicate correlations in the Hubbard leg and filled symbols in the Fermi leg.}
\label{fig:pdwcordoped}
\end{figure}

PDW correlations in the lightly doped spin-gapped Mott phase exhibit a power law with an exponent close to -2 as
depicted in Fig.~\ref{fig:pdwcordoped}. Thus they are not (or barely) enhanced in comparison to
a noninteracting ladder ($U=0$) and cannot explain the pair binding observed in the excitation energies and local densities of that phase~\cite{abd15}.
Density-density correlations (not shown) decrease as fast as the PDW correlations.
The dominant correlations seem to be pow-law SDW correlations with exponents close to {-1}, which are also shown in Fig.~\ref{fig:pdwcordoped}.
Here, we cannot decide whether this power-law behavior is real (as in the Luttinger liquid phase) or a finite-size effect (as in the Kondo-Mott insulator)
because we could not perform an accurate finite-size scaling of the spin gap with the accessible ladder lengths (see Sec.~\ref{sec:extrap}).
Enhanced PDW correlations were found in doped Kondo-Heisenberg ladders with a substantial spin gap~\cite{ber10}.
These results suggest a competition between the SDW and PDW fluctuations in asymmetric ladders.
Note that the analysis of correlation functions away from half filling is a delicate problem because of the
inhomogeneous distribution of charge and spin along the ladder~\cite{abd15} and between both legs (see the next section).

In summary, this investigation of correlation functions is compatible with the phase diagram deduced in our previous work~\cite{abd15}.
The only discrepancy is the apparent power-law behavior of spin correlations in the spin-gapped Kondo-Mott phase, which we can 
understand as a finite-size effect but should be checked using longer ladder lengths in a future study.
In the Luttinger liquid phase, charge and spin fluctuations appear to be spatially separated with the stronger spin fluctuations
in the Hubbard leg and the stronger charge fluctuations in the Fermi leg.
Despite the strong pair binding in the Fermi leg of the (doped) spin-gapped Mott phase, which we deduced from the energy and density observables, 
we could not identify any enhanced pairing correlations and the dominant correlations seem to be SDW in that regime.
Therefore, we think that the coexistence of (quasi-)long range orders is likely
in systems of coupled asymmetric Hubbard ladders or in two-dimensional systems made of a Hubbard layer and a Fermi
layer.
We expect antiferromagnetic spin order in the interacting Hubbard subsystem while various pairing or charge orders
could dominate the noninteracting Fermi subsystem.

\section{Leg densities}\label{sec:dens}

We can obtain interesting information about the low-energy excitations of the asymmetric ladder~(\ref{eq:hamiltonian}) 
using the changes in the total charge and spin densities on both legs for variable numbers of electrons in the system.
The deviations from the ground-state charge and spin distributions at half filling are given by
\begin{equation}
 N_{m}(y)= \sum_{x}\langle N(x,y) \rangle -L_x
 \label{eq:charge}
\end{equation}
and
\begin{equation}
 S_{m}(y)= \sum_{x} \langle S(x,y) \rangle .
 \label{eq:spin}
\end{equation}
Here, we will consider the deviations caused by one ($m$=1p) or two electrons ($m=$2p) added and by a spin triplet ($m$=1s).
This corresponds to the lowest single-particle, charge and spin excitations, respectively (see the discussion of gaps below).

\begin{figure}
\centering
\includegraphics[width=0.45\textwidth]{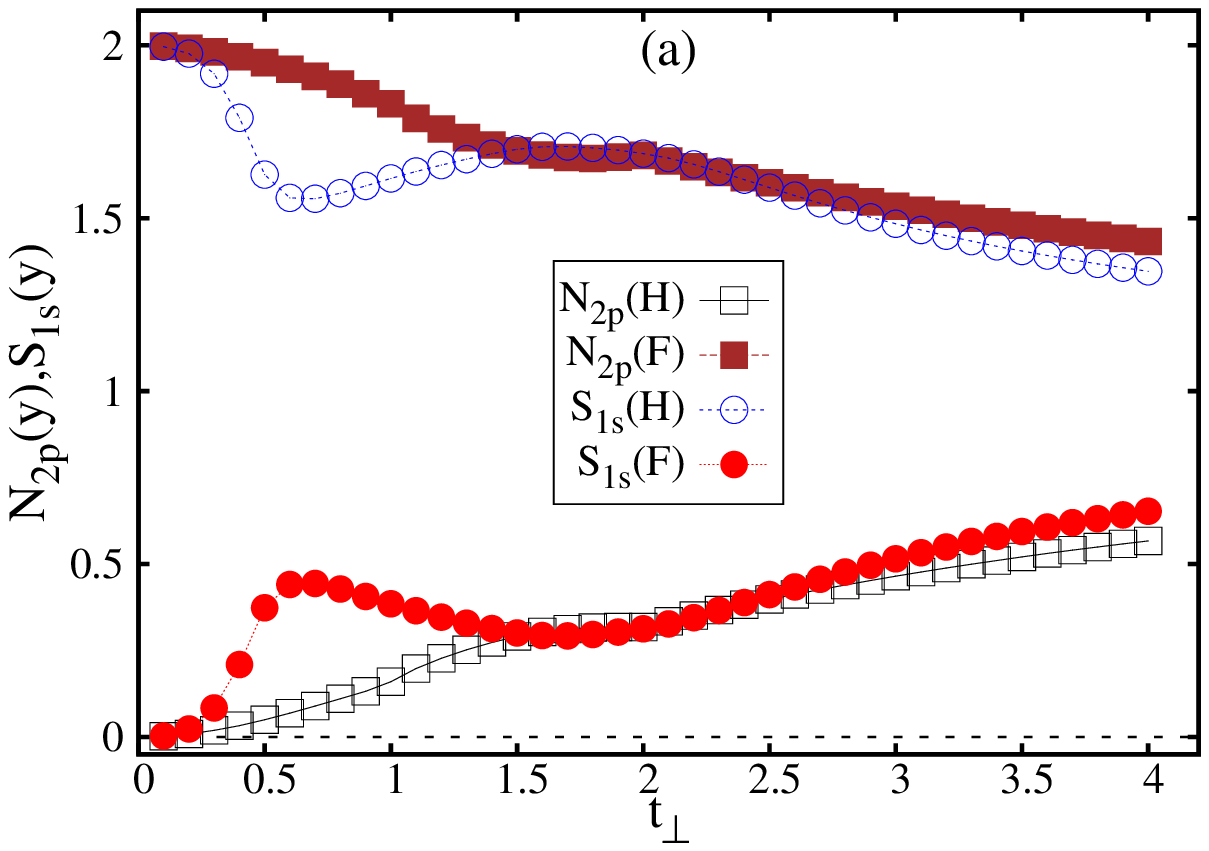}
\includegraphics[width=0.45\textwidth]{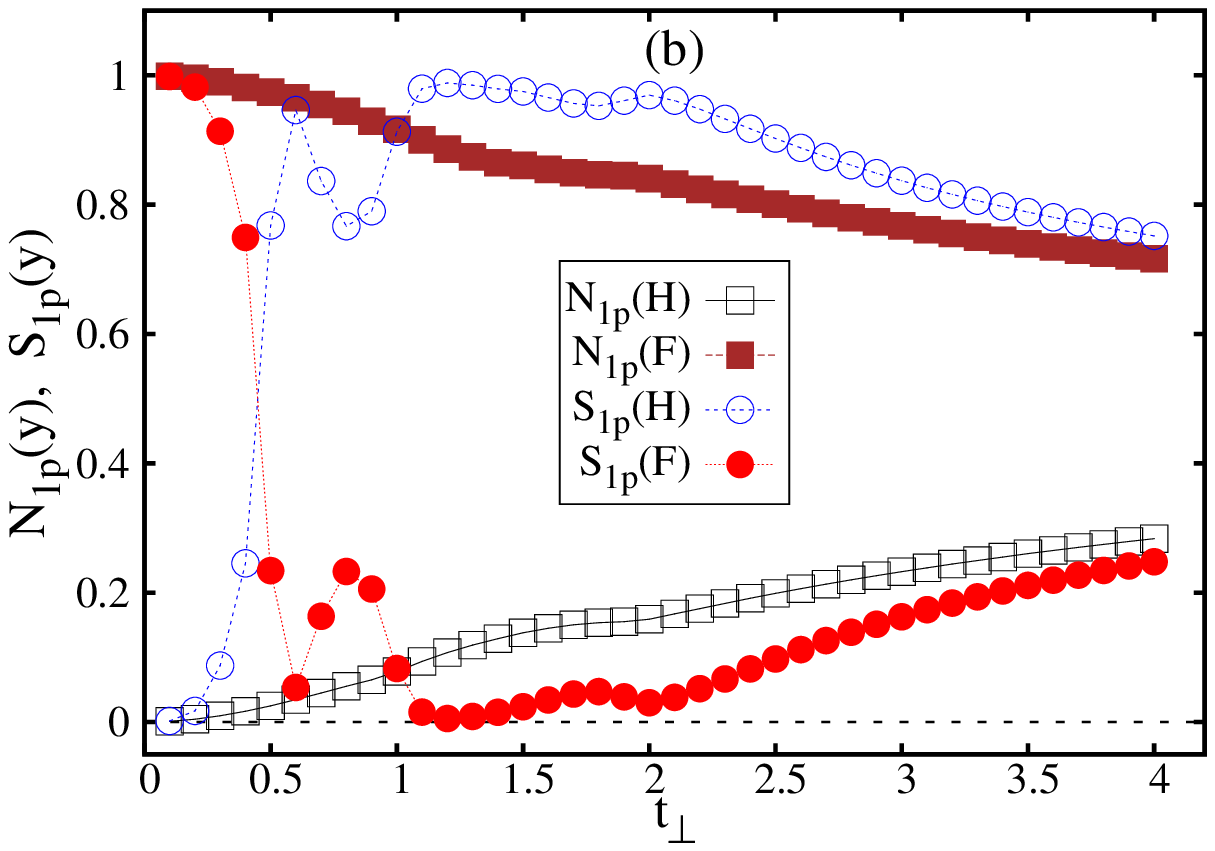}
\caption{Deviations of the total charge~(\ref{eq:charge}) and spin~(\ref{eq:spin}) on the Hubbard leg 
with $U=8$
(open symbols) and the Fermi leg (solid symbols)
as a function of the inter-leg hopping $t_{\perp}$: 
(a) $N_{\text{2p}}(y)$ for two added electrons (squares) and $S_{\text{1s}}(y)$ for one spin triplet excitation (circles).
(b) $N_{\text{1p}}(y)$ (squares) and $S_{\text{1p}}(y)$ (circles) for one added electron.
}
\label{fig:chrgspindens}
\end{figure}

Figure~\ref{fig:chrgspindens}(a) shows $N_{\text{2p}}(y)$ and $S_{\text{1s}}(y)$ for the lowest charge and spin excitation as a function
of $t_{\perp}$. 
Clearly, most of the excess charge is concentrated on the Fermi leg
while most of the excess spin is localized on the Hubbard leg. 
The confinement of additional charges in the Fermi leg is expected because of the repulsive interaction on the Hubbard leg.
The confinement of the excess spin in the Hubbard leg is more surprising because both legs have gapless spin excitations
when decoupled ($t_{\perp}=0$). 
This uneven distribution is probably related to the fact that spin excitations have a lower
velocity in the 1D Hubbard model for $U>0$ than in the tight-biding chain~\cite{hubbard-book}.
Thus the lowest charge and spin excitations are separated in real space in this model.
In the Luttinger liquid phase, we see that this separation is almost perfect with most
of the density variations concentrated in opposite legs. We will showd in the next section that the lowest excitations
are also dynamically separated, i.e. have different velocities, in that phase.
For increasing coupling $t_{\perp}$, the distributions of charge and spin become progressively more even
and converge to the same values for both legs in the dimer limit ($t_{\perp} \gg t_{\parallel}, U$).
However, a  non-monotonic behavior of the spin distribution $S_{\text{1s}}(y)$ is observed for hopping terms $t_{\perp}$
corresponding to the Kondo-Mott and spin-gapped Mott phases. 

The picture is somewhat different for the single-particle excitation in Fig.~\ref{fig:chrgspindens}(b).
Both the additional charge $N_{\text{1p}}(y)$ and spin $S_{\text{1p}}(y)$  are localized in the Fermi leg in the Luttinger liquid phase. 
However, the excess spin moves onto the Hubbard leg upon entering the Kondo-Mott phase with increasing $t_{\perp}$.
Then the density deviations for the single electron behave similarly to those for charge and spin excitations
in Fig.~\ref{fig:chrgspindens}(a): overall convergence toward equal values in the dimer limit and non-monotonic behavior
in  the Kondo-Mott and spin-gapped Mott phases.

In summary, the low-energy excitations are mostly confined to one leg when the rung hopping $t_{\perp}$ is not too strong.
This explains the different behavior of correlation functions on Hubbard and Fermi legs.
In most cases charge excitations tend to stay on the Fermi leg, while spin excitations prefer the Hubbard leg.
Thus low-energy spin and charge excitations are spatially separated in the 1D correlated electron model~(\ref{eq:hamiltonian}).
The single-particle excitations in the Luttinger liquid phase constitute the only exception.
The difference between pure spin excitations and the spin associated to single-particle excitations in that phase is intriguing
and we have investigate the velocities of these excitations to gain more information.

\section{Velocities}\label{sec:extrap}

Similarly to the analysis of the leg density distributions for excited states~(\ref{eq:charge}) and~(\ref{eq:spin}),
we calculate the gaps for single-particle, charge, spin and excitations from the change in ground-state energies
for one added particle (electron or hole), two added particles (electrons or holes) and a spin triplet.
The charge gap is defined as
\begin{eqnarray}
E_{\text{c}} & = & \frac{1}{2} 
\left [ E_{0}(N_{\uparrow}+1,N_{\downarrow}+1)+E_{0}
(N_{\uparrow}-1,N_{\downarrow}-1) \right . 
\nonumber \\ 
&& \left . -2E_{0}(N_{\uparrow},N_{\downarrow})) \right ]
\label{eq:chargegap}
\end{eqnarray}
where $E_{0}(N_{\uparrow},N_{\downarrow})$ refers to the ground-state
energy of the Hamiltonian~(\ref{eq:hamiltonian}) for $N_{\sigma}$ electrons of spin $\sigma$.
This gap is the lowest excitation energy seen in the dynamical structure factor, which can be measured in experiments such as electron energy loss spectroscopy.
The spin gap is defined as
\begin{equation}
E_{\text{s}}=E_{0}(N_{\uparrow}+1,N_{\downarrow}-1)-E_{0}(N_{\uparrow},N_{\downarrow}) .
\label{eq:spingap}
\end{equation}
This gap is the lowest excitation energy seen in the the dynamical spin structure factor, which can be measured in experiments such as inelastic neutron-scattering.  
Finally, the single-particle gap is defined as
\begin{equation}
E_{\text{p}}=E_{0}(N_{\uparrow}+1,N_{\downarrow})+
E_{0}(N_{\uparrow}-1,N_{\downarrow})-2E_{0}(N_{\uparrow},N_{\downarrow}).
\label{eq:particlegap}
\end{equation}
This gap is the lowest excitation energy seen in the single-particle spectral functions (Green's functions), which can be probed in experiments such as 
angle resolved photoemission spectroscopy.

In a ladder~(\ref{eq:hamiltonian}) of finite length $L$, these gaps are always finite (excluding accidental degeneracies).
To determine the true gaps of an infinite ladder, one has to analyze the scaling  of the finite-size gaps with the ladder length.
The finite-size scaling is performed by calculating these gaps for several system sizes up to $L=200$ using DMRG and 
extrapolating the values to $L \rightarrow \infty$ numerically. 
For most of the parameter space $(U, t_{\perp})$ of the half-filled Hamiltonian~(\ref{eq:hamiltonian}) we found that
the gaps remain finite in the thermodynamic limit. This corresponds to the three insulating phases
in the phase diagram in Fig.~\ref{fig:model}. These results were presented in detail in our previous work~\cite{abd15}.
Here, we focus on the gapless Luttinger liquid phase.

\begin{figure}
\centering
\includegraphics[width=0.45\textwidth]{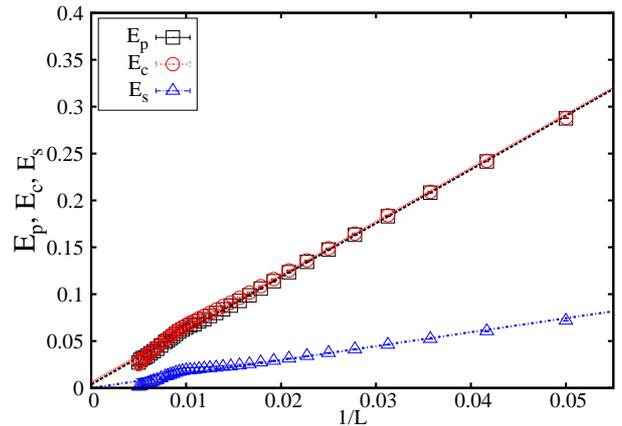}
\caption{Charge gap ($E_{\text{c}}$), single-particle gap ($E_{\text{p}}$),
and spin gap ($E_{\text{s}}$) of the half-filled asymmetric Hubbard ladder~(\ref{eq:hamiltonian}) 
as a function of the inverse ladder length  $1/L$ for $U=8$ and $t_{\perp}=0.3$.}
\label{fig:EcEpEsgappless}
\end{figure}

Figure~\ref{fig:EcEpEsgappless} shows that
the extrapolation of the finite-size gaps
 indicate gapless excitations in the thermodynamic limit for $U=8$ and $t_{\perp}=0.3$, which corresponds to
the Luttinger liquid phase. Moreover, the gaps tend to vanish linearly with the inverse system length as expected for
1D correlated conductors~\cite{giamarchi}. The slope corresponds to the excitation velocity up to a constant prefactor $\pi$
(assuming $\hbar=1$ and a lattice constant $a=1$).
The deviations from the linear behavior for small $1/L$ are due to
rapidly increasing relative errors because the absolute DMRG errors for the energies $E_{0}(N_{\uparrow},N_{\downarrow})$ 
scale as $L$ while the energy differences~(\ref{eq:chargegap}),~(\ref{eq:spingap}), and~(\ref{eq:particlegap})
are of the order of $1/L$.

The finite-size charge gap scales as $E_{\text{c}}\approx 5.8/L$ for large ladder length $L$,
in agreement with the finite-size scaling in the half-filled tight-binding chain with open boundary conditions, 
i.e. $E_{\text{c}}\approx 2\pi/L$.
This confirms that the added charges are concentrated mostly on the Fermi leg.
Similarly, the finite-size spin gap scales as $E_{\text{s}}\approx 1.49/L$
in agreement with the Bethe Ansatz solution for the one-dimensional Hubbard model with $U=8$ and open boundary conditions,
which yields $E_{\text{s}}\approx 1.51/L$~\cite{hubbard-book}. 
This also confirms the localization of the lowest triplet excitation in the Hubbard leg.

The finite-size single-particle gap is very close to the charge gap as seen in Fig.~\ref{fig:EcEpEsgappless}. 
In a Luttinger liquid the single-particle velocity is the average of the velocities for the elementary charge and spin excitations
that contribute to the lowest single-particle excitations. Consequently, the elementary spin excitation contributing to
the single-parti\-cle excitation in Fig.~\ref{fig:EcEpEsgappless} must have almost the same velocity than the elementary
charge excitation and thus is not responsible for the finite-size spin gap seen in that figure.

Therefore, the nature of elementary excitations in the Luttinger liquid phase is relatively simple.
In an electronic two-leg ladder, two charge modes and two spin modes can exist~\cite{giamarchi}.
We have seen in the previous section that, in the Luttinger liquid phase of the asymmetric Hubbard ladder, 
each mode is concentrated in one leg.
The charge mode in the Hubbard leg is gapped and thus not relevant for the Luttinger liquid properties.
The spin mode in the Hubbard leg is gapless, determines the finite-size spin gap seen in Fig.~\ref{fig:EcEpEsgappless} for triplet excitations,
and is responsible for the critical antiferromagnetic correlations that can be seen in Fig.~\ref{fig:szszcor}(a)
and the spin density deviations caused by the triplet excitation in Fig.~\ref{fig:chrgspindens}(a).
The low-energy single-particle excitations are essentially made of charge and spin excitations localized in the Fermi leg.
The spin mode in the Fermi leg is gapless but has a higher velocity and thus larger finite-size gaps than 
the spin mode in the Hubbard leg.
It is responsible for the weak antiferromagnetic correlations in the Fermi leg that can be seen in Fig.~\ref{fig:szszcor}(a)
and the spin density deviations caused by the single-particle excitation in Fig.~\ref{fig:chrgspindens}(b).
Finally, the charge mode in the Fermi leg is gapless, has approximately the same velocity than the spin mode and is responsible
for the charge density deviations seen in Fig.~\ref{fig:chrgspindens}(a) and~(b), as well as the power-law  density correlations
in Fig.~\ref{fig:denscor}(a).
The near equality of the spin and charge correlations and velocities in the Fermi leg suggest that the effective Luttinger liquid
induced in this leg is only weakly correlated.

\section{Conclusions}\label{sec:concl}

We have investigated the four ground-state phases found previously~\cite{abd15} 
in the half-filled asymmetric Hubbard ladder using the DMRG method.
The correlation functions studied in Sec.~\ref{sec:corr} are fully compatible with our previous findings,
besides the problem of apparent power-law correlations in phases with very small spin gaps.
Quasi-long-range antiferromagnetic order is found only in the Hubbard leg of the Luttinger liquid phase
and, possibly, upon doping of the spin-gapped Mott phase.
An open issue is the absence of enhanced pairing correlations despite the strong pair binding observed
in the excitation energies and local densities in the Fermi leg of the (lightly doped) spin-gapped Mott phase.

The leg density distributions discussed in Sec.~\ref{sec:dens} confirm the existence and the (rough) location 
of the four ground-state phases. 
So far our investigations have not yield precise phase boundaries
(and consequently no information on the nature of the phase transitions), except in limiting cases.
Entanglement measurements~\cite{ost02,wu04,leg06,leg07,mun09} based on the DMRG method, 
such as the block entanglement entropy, are the most promising approach 
to determine these phase boundaries and are in progress.

Correlation functions, leg densities, and excitation velocities show that low-energy spin and charge degrees 
of freedom can be spatially separated in the asymmetric Hubbard ladder.
This confinement results in different correlations and low-energy excitations in both legs
in some cases. Thus it   
could facilitate the coexistence of various (quasi-)long-range orders
in the Hubbard and Fermi subsystem of generalizations of the asymmetric Hubbard ladder.
In particular, the Hamiltonian~(\ref{eq:hamiltonian}) can be generalized to allow
for different intra-leg hoping terms in both legs and thus 
to reach the regime of 
strong spin-exchange coupling on the rung (compared to the hopping in the Fermi leg), 
where enhanced PDW correlations were found in the Kondo-Heisenberg model~\cite{ber10}.  
Therefore, we think that the coexistence of (quasi-)long range orders could be possible  
in two-dimensional models made of coupled (generalized) asymmetric Hubbard ladders 
or of a Hubbard layer and a Fermi layer.
Antiferromagnetic spin order should occur in the interacting Hubbard subsystem while 
pairing or other charge orders could dominate the noninteracting Fermi subsystem.
These models could describe real quasi-two-dimensional materials such as the 
layered high-temperature superconductors or arrays of linear atomic chains deposited on
semiconducting substrates~\cite{erw10,aul13}.

\begin{acknowledgement}
This work was done as part of the Research Units \textit{Metallic nanowires on the atomic scale: Electronic
and vibrational coupling in real world systems} (FOR1700) of the German Research Foundation (DFG) and was supported by
grant~JE~261/1-2. The DMRG calculations were carried out on the cluster system
at the Leibniz University of Hannover and at the Sudan Center for HPC and Grid Computing.
\end{acknowledgement}

\section*{Author contribution statement}
Eric Jeckelmann initiated and guided the project.
Anas Abdelwahab adapted the DMRG code and performed the DMRG simulations.
Both authors analyzed the data and contributed to the writing of the manuscript.

\end{document}